\documentclass[reprint,amsmath,amssymb,aps,showkeys]{revtex4-2}

\usepackage{slashed}
\usepackage{amsmath}

\usepackage{graphicx}
\usepackage{dcolumn}
\usepackage{bm}
\usepackage{hyperref}
\usepackage[mathlines]{lineno}
\usepackage{xcolor}
\usepackage[makeroom]{cancel}

\usepackage{upgreek}

\newcommand\N{\mathcal{N}}

\renewcommand\H{\mathcal{H}}

\newcommand\A{\mathcal{A}}

\newcommand\F{\mathcal{F}}

\newcommand\K{\mathcal{K}}
\newcommand\J{\mathcal{J}}

\newcommand\D{\mathcal{D}}

\newcommand\n{\text{\tiny{N}}}

\newcommand\rarrow{\rightarrow}

\newcommand\LieK{\mathfrak{K}}

\renewcommand\b{\bar }

\renewcommand\d{\partial}

\newcommand\bs{\boldsymbol}

\renewcommand\-{^{-1}}

\DeclareMathOperator{\Diff}{Diff}

\begin{document}


\title{Off-shell supersymmetry via manifest invariance}

\author{J. François}
\email{jordan.francois@uni-graz.at}
\affiliation{University of Graz (Uni Graz), 
Heinrichstraße 26/5, 8010 Graz, Austria, and \\
Masaryk University (MUNI), 
Kotlářská 267/2, Veveří, Brno, Czech Republic, and \\
Mons University (UMONS), 
20 Place du Parc, 7000 Mons, Belgium. 
}

\author{L. Ravera}
\email{lucrezia.ravera@polito.it}
\affiliation{Politecnico di Torino (PoliTo),
C.so Duca degli Abruzzi 24, 10129 Torino, Italy, and \\
Istituto Nazionale di Fisica Nucleare (INFN), Section of Torino,
Via P. Giuria 1, 10125 Torino, Italy, and \\
Grupo de Investigación en Física Teórica (GIFT),
Universidad Cat\'{o}lica De La Sant\'{i}sima Concepci\'{o}n, Concepción, Chile.
}

\date{\today}

\begin{abstract}

A fundamental challenge in supersymmetric field theory is that supersymmetry transformations on field variables generally form an algebra only on-shell, i.e. upon imposing the field equations. 
We show that this issue is defused in a manifestly relational -- and thus automatically invariant -- formulation of supersymmetric field theory, achieved through the application of the Dressing Field Method of symmetry reduction, a 
systematic tool to exhibit the gauge-invariant content of general-relativistic gauge field theories. 

\end{abstract}


\maketitle


\section{Introduction}\label{Introduction}

Finding a general \emph{off-shell} formulation of supersymmetric field theories, a 
very desirable goal 
aiming
towards 
their description as quantum field theories, is a major challenge of theoretical physics of fundamental interactions. 
For theories with more than eight supercharges such a formulation is typically problematic,
both in the rigid supersymmetry (susy) and in the supergravity (sugra) cases. 
The difficulties are magnified when coupling $\mathcal{N}$-extended
theories to hypermultiplets, i.e.
matter multiplets with maximum spin 1/2 and whose number of d.o.f. remains
stable when moving from massive to massless representations. An off-shell supersymmetric description in ordinary
superspace $M^{D|\n \mathcal{N}}$, with $\n=2^{D/2}$ and $D$ the number of spacetime dimensions, can be formulated only for a very restricted class of
supermultiplets \cite{deWit:1979xpv,Fradkin:1979cw,Fradkin:1979as,deWit:1980lyi,Howe:1982tm,deWit:1982na,Galperin:1986fg,Gonzalez-Rey:1997pxs,deWit:2006gn,Gates:2014vxa}. 
A general off-shell description of representations for $\mathcal{N}$-extended models, with fully
manifest susy, was shown to be possible at the cost of including an infinite number of auxiliary fields, thereby circumventing the no-go theorems of susy stating that adding any finite
number of auxiliary fields 
cannot
help to
achieve
the \emph{closure of the off-shell algebra} \cite{Howe:1985ar,Rivelles:1982gn,Bazhanov:1980ku}. 
The most successful frameworks in which
off-shell theories have been technically developed involving an infinite number of auxiliary fields are \emph{projective superspace} \cite{Karlhede:1984vr,Lindstrom:1989ne} and \emph{harmonic superspace} \cite{Galperin:2001seg}. {Another powerful approach to off-shell susy considers, instead, constrained ``pure" spinors -- see \cite{Cederwall:2013vba,Cederwall:2022fwu} and references therein.} However, the physical and group-theoretical meaning of
having an infinite number of auxiliary fields {or constrained fields} remains unclear.

It is important to recall what is expected, within the standard supersymmetric field theory framework, regarding a general off-shell formulation. 
Such a  formulation is indeed structured around two fundamental criteria:
\begin{enumerate}
    \item[(i)] The off-shell matching of bosonic and fermionic degrees of freedom (d.o.f.), which must also reappear on-shell. However, this matching is initially computed only at the kinematical level, taking into account the gauge transformations of the fields -- typically -- or any constraints, depending on the approach adopted.
    \item[(ii)] The closure of an off-shell algebra for the susy transformations acting on the fields. This means that the fields form an off-shell representation of the superalgebra, or the superalgebra is represented off-shell on the fields 
    -- this is the so-called off-shell closure.
\end{enumerate}

Typically, in the absence of auxiliary fields, the ``algebra" of susy transformations closes only on-shell on field variables, i.e. only when the equations of motion are used. This implies that susy transformations do not actually form a true algebra but rather what is sometimes referred to as an \emph{open} or \emph{soft} algebra \cite{Galperin:2001seg}. 

The above two points should be emphasized separately, as in some cases it is possible to construct off-shell supermultiplets even with a \emph{finite} number of fields, ensuring the matching of bosonic and fermionic degrees of freedom. However, the susy transformations do not close off-shell on the fields, they only close on-shell. Consequently, in such cases, one cannot speak of a manifestly off-shell supersymmetric theory.
For example, as noted in \cite{Brandt:2000uw} and more recently revisited in \cite{Andrianopoli:2024ynz}, this occurs in the case of the so-called \emph{double tensor multiplets} \cite{Theis:2002er,Theis:2003jj,Djeghloul:2012pr}. Similarly in $\mathcal{N}$-extended models and in higher-dimensional theories, see e.g. \cite{DAuria:1984glj}.

In this letter, we do not aim to provide a definitive solution to both of the aforementioned issues simultaneously. Instead, we focus on (ii), by proposing a novel alternative formulation of standard supersymmetric field theory with a \emph{finite} number of fields, in which susy is reduced using the \emph{Dressing Field Method} (DFM). 
Originally introduced in \cite{Francois2014} as a \emph{systematic approach} to constructing gauge-invariant variables, the DFM is best framed within the differential bundle geometry of field space \cite{JTF-Ravera2024gRGFT}, but it also admits a more broadly accessible field-theoretic presentation \cite{JTF-Ravera2024-SUSY, JTF-Ravera2025DFMSusyReview&Misu, Berghofer-Francois2024}. 
A key aspect of this method, particularly when applied using \emph{field-dependent dressing fields}, is its \emph{relational} nature: gauge invariance is achieved by realizing the \emph{physical} d.o.f. as \emph{relations} among bare (gauge-variant) ones \cite{Francois2023-a,JTF-Ravera2024gRGFT,JTF-Ravera2024-SUSY,JTF-Ravera2024ususyDFM,JTF-Ravera2024NRrelQM,JTF-Ravera2025DFMSusyReview&Misu}.

By leveraging this approach we make invariance manifest through the construction of relational susy singlets, thereby defusing the issue of susy transformations closing and forming an algebra only on-shell.

The remaining of this work is thus organized as follows:
In Section \eqref{Dressing Field Method: basics} we briefly recall the basics of the DFM and its application to the reduction of susy -- while we refer the interested reader to the above cited literature on the DFM for further technical and conceptual details. {This more formal section provides the general mathematical template of the DFM, to better appreciate its field-theoretic application to the susy case. A more physics-oriented reader may start directly with Section \ref{Off-shell susy via manifest invariance}, where} we present our novel approach to off-shell susy invariance via the DFM. The procedure applies at both the kinematical and the dynamical levels. We also provide an  example of application to an $\mathcal{N}$-extended model, that is $\mathcal{N}=2$, $D=4$ pure sugra in its geometric (a.k.a. \emph{rheonomic}) formulation in superspace \cite{Castellani:1991eu} -- the dressing will be perturbative and invariance achieved at 1st order. 
Finally, in Section \ref{Discussion} we
discuss future developments and propose novel possible approaches to addressing point (i) using a finite number of fields.

\section{Dressing Field Method: basics}\label{Dressing Field Method: basics}

Via the DFM \cite{GaugeInvCompFields,Berghofer-et-al2023,Francois2023-a,JTF-Ravera2024gRGFT,JTF-Ravera2024-SUSY,JTF-Ravera2024ususyDFM,JTF-Ravera2024NRrelQM,JTF-Ravera2025DFMSusyReview&Misu} one produces gauge-invariants out of the fields  $\Phi=\{\upphi\}$ of a gauge theory with gauge group $\H$ whose action on $\Phi$ defines gauge transformations: $\upphi \mapsto \upphi^{g}$.
In the standard formulation of the DFM, which we review below, one first studies the kinematics (and then the dynamics) of a theory based on a given gauge group, for which the transformations that form a closed group, or an algebra, are given a priori. 

\smallskip

Suppose $\K \subseteq \H$ is a gauge subgroup -- corresponding to the rigid $K\subseteq H$. The DFM relies on identifying a \emph{$\Phi$-dependent $\K$-dressing field}, that is a map
\begin{equation}
\label{Field-dep-dressing}
\begin{aligned}
u :  \Phi \ &\rarrow\  \D r[K, \K], \\
    \upphi \  &\mapsto\  u=u[\upphi], \\
    \upphi^k &\mapsto\ u^k:=u[\upphi^k]=k\-u[\upphi], \ \ \forall k\in \K,
\end{aligned} 
\end{equation}
where $\D r[K, \K]:=\{u: U\subset M \rarrow K\,|\, u^k=k\-u\}$ is the space of ($\Phi$-independent) dressing fields, with $M$ the spacetime manifold.
One can then build in a systematic way, using the DFM rule of thumb -- cf. e.g. \cite{JTF-Ravera2024gRGFT}, the $\K$-\emph{invariant dressed fields}
by the surjective map
\begin{equation}
\label{Dressed-field}
\begin{aligned}
 \Phi \ &\rarrow\  \Phi^u, \\
    \upphi \  &\mapsto\  \upphi^u=\upphi^{u[\upphi]}, \\
    \upphi^k &\mapsto\ (\upphi^k)^{u^k}:=(\upphi^k)^{k\-u[\upphi]}=\upphi^{u[\upphi]}.
\end{aligned} 
\end{equation} 
By this rule, to obtain the dressing of an object, one first computes  its gauge transformation, then one (formally) substitutes the gauge parameter with $u$ in the result. Note, however, that the dressing field \emph{is not} an element of the gauge group.
The DFM has a natural relational interpretation: Dressed fields $\upphi^u$ represent the gauge-invariant, physical \emph{relations} among  d.o.f. embedded in the original (bare) fields $\upphi$.

\smallskip

Here let us also mention that, being $\K$-invariant, the dressed fields $\upphi^u$ are expected to display \emph{residual transformations} under what remains of the gauge group. 
If $K$ is a normal subgroup of $H$, $K \triangleleft H$, then $H/K=: J$ is a Lie group. 
Correspondingly, $\K \triangleleft \H$ and $\J =\H/\K$ is a gauge subgroup of $\H$.
In this case, the dressed fields $\upphi^u$ may exhibit well-defined residual $\J$-gauge transformations, which are called \emph{residual transformations of the 1st kind}. Furthermore, dressed objects may also exhibit residual transformations resulting from a possible ambiguity in the choice of dressing field: two dressing fields $u$, $u'$ may a priori be related by $u'=u\xi$, where $\xi$ is an element of what is referred to as the group of \emph{residual transformations of the 2nd kind}.
\smallskip

{
Let us observe that 
if a dressing is introduced by \emph{fiat}, as additional d.o.f. not built from those of $\Phi$, i.e. it is a $\Phi$-independent dressing field,
one is dealing with a new \emph{distinct} theory: 
The dressing field is said \emph{ad hoc}, the dressed fields cannot be interpreted as representing the physical relational content of the original theory, and
the gauge symmetry of the new theory is said to be \emph{artificial} \cite{Pitts2009,Francois2018}.
This encompasses as a special case the so-called Stueckelberg trick, whereby one \emph{implements} a gauge symmetry in a theory via the introduction of extra d.o.f., the Stueckelberg fields, which are indeed \emph{ad hoc} dressing fields. 
Looking at  Stueckelberg models (like massive Yang-Mills and massive gravity) through the DFM lens, one would ``undo" the result of the Stueckelberg trick, removing an artificial gauge symmetry.}

\subsection{Perturbative dressing}

It may often happen, as is typically the case in supersymmetric field theory, that one is chiefly concerned with \emph{invariance at 1st order}, that is under infinitesimal gauge transformations Lie$\H$. In this case, one may linearise the above, defining a Lie$\K$-dressing field
\begin{equation}
\label{dressingfieldtr}
\begin{aligned}
&\upsilon=\upsilon[\upphi]: U\subset M \rarrow \LieK=\text{Lie}K, \\[1mm]
& \text{s.t. } \quad \delta_\lambda \upsilon :=\upsilon[\delta_\lambda \upphi] \approx -\lambda , \ \ \forall\lambda \in \text{Lie}\K,
\end{aligned}
\end{equation}
where in the defining transformation higher-order terms, polynomial in $\lambda$ and $\upsilon$, are neglected.
We then define the \emph{perturbatively dressed fields} 
\begin{align}
\label{pert-dressed-fields}
\upphi^\upsilon:= \upphi + \b\updelta_\upsilon \upphi,
\end{align}
where $\b\updelta_\upsilon \upphi$ mimics the functional expression of the Lie$\H$ gauge transformation $\delta_\lambda \upphi$, substituting the gauge parameter by the infinitesimal dressing, $\lambda \rarrow \upsilon$. 
This is again the rule of thumb mentioned above.
So,  $\b\updelta_\upsilon$ \emph{is not} a differential of the algebra of fields.
The perturbatively dressed fields are $\K$-invariant at 1st order,
\begin{equation}
\begin{aligned}
\delta_\lambda (\upphi^\upsilon)
&= \delta_\lambda \upphi + \b\updelta_{(\delta_\lambda \upsilon)} \upphi 
= \delta_\lambda \upphi + \b\updelta_{-\lambda} \upphi \\
& = \delta_\lambda \upphi - \delta_\lambda \upphi \equiv 0,
\end{aligned}
\end{equation}
that is neglecting higher-order terms in $\lambda$ and $\upsilon$. 

\subsection{DFM and supersymmetry reduction}

For supersymmetric field theories the reasoning is slightly different from the standard gauge theoretic framework  where the DFM naturally applies, in the sense that one usually starts from a Lagrangian functional of the fields -- hence from some dynamical data -- required to be (quasi-)invariant (i.e.,  invariant up to a
boundary term) under susy transformations that generally close an algebra only on-shell. Nonetheless, the DFM can still be directly applied to this  context, reducing susy and building susy-invariant fields.
Unfortunately, in field-theoretic contexts dressings are often conflated with mere gauge-fixings, generating misconceptions regarding the physics -- cf. \cite{JTF-Ravera2024ususyDFM,JTF-Ravera2025DFMSusyReview&Misu,Berghofer-Francois2024} for details.

As far as standard  susy  (as opposed to unconventional susy \cite{JTF-Ravera2025DFMSusyReview&Misu}) is concerned, it was shown in \cite{JTF-Ravera2024-SUSY} that the ``gauge-fixing" conditions typically used to extract the d.o.f. of the Rarita-Schwinger (RS) spinor-vector and gravitino fields -- e.g., 12 off-shell d.o.f. in $\mathcal{N}=1$, $D=4$ -- are actually instances of the DFM. That is, solving the dressing functional constraints actually realises those fields as (self-dressed) relational variables. 
In both cases of rigid susy and sugra, what is commonly imposed is the so-called gamma-tracelessness condition on the fermionic 1-form $\psi^\alpha=\psi^\alpha_\mu dx^{\,\mu}$, namely $\gamma^\mu \psi_\mu = 0$ (for simplicity, from now on we will omit all spinor indices). 
This condition is rewritten for the new dressed variable, i.e. $\gamma^\mu \psi^u_\mu=0$, obtained by applying the DFM rule of thumb, 
and then explicitly solved for $u$, showing that it indeed qualifies as a dressing field after verifying how it transforms under susy. In this context, the dressing is \emph{field-dependent} and \emph{non-local} -- and thus susy a substantive symmetry \cite{Francois2018}.
In the RS case, in $D$ spacetime dimensions, {we have that the RS field $\psi$ transforms under susy as follows:
\begin{align}
    \psi \mapsto \psi + d \varepsilon ,
\end{align}
$\varepsilon$ being the susy parameter. Then, one introduces the (dressed) variable
\begin{align}
    \psi^u := \psi + d u , \quad \text{that is} \quad \psi^u_\mu := \psi_\mu + \partial_\mu u ,
\end{align}
and, solving the ``dressing constraint" $\gamma^{\mu} \psi^u_\mu=0$ for $u$, one gets}
\begin{equation}
\begin{aligned}
    \label{gamma-tr-dressing}
    & \gamma^{\mu} \psi^u_\mu = \gamma^{\mu} (\psi_\mu + \d_\mu u) = 0 \\
    & \Rightarrow \quad u[\psi] = - \slashed{\d}\- (\gamma^{\mu} \psi_\mu) = - D \slashed{\d}\- \chi,
\end{aligned}
\end{equation}
where $\chi:=1/D \, \gamma^{\mu} \, {\psi}_{\mu}$ carries spin-1/2. {Notice that $\chi^u=0$ by construction. See \cite{JTF-Ravera2024-SUSY} for details.}
The minimal coupling of the RS field with gravity is
described via the Lorentz spin connection $\omega^{ab}$ and the soldering 1-form $e^a$ (the vierbein). It is obtained via covariantization,
\begin{equation}
\begin{aligned}
    \d_\mu & \,\mapsto\, \D_\mu , \\
    \gamma_\mu & \,\mapsto\, \gamma_a {e^a}_\mu ,
\end{aligned}
\end{equation}
where $\D_\mu$ is the Lorentz-covariant derivative and $\gamma_a$ the flat space gamma-matrices. 
Then, in the simple $\mathcal{N}=1$, $D=4$ sugra case, we consider the gamma-tracelessness constraint as a functional condition on 
\begin{align}
    \psi_\mu^\upsilon := \psi_\mu+ \b\updelta_\upsilon \psi ,
\end{align}
with $\psi$ the gravitino field.
Solving explicitly for $\upsilon$, one obtains the field-dependent \emph{perturbative dressing field} $\upsilon$:
\begin{equation}
\begin{aligned}
\label{gamma-tr-dressing-sugra}
    & \gamma^{\mu} \psi^\upsilon_\mu = \gamma^{\mu} (\psi_\mu + \D_\mu \upsilon) = 0 \\
    & \Rightarrow \quad \upsilon[\psi] = - \slashed{\D}\- (\gamma^{\mu} \psi_\mu) = - D \slashed{\D}\- \chi.
\end{aligned}
\end{equation}
One shall then dress also the other fields of the theory, if any, as well as  the Lagrangian -- see \cite{JTF-Ravera2024-SUSY,JTF-Ravera2025DFMSusyReview&Misu}.

\section{Off-shell susy via manifest invariance}\label{Off-shell susy via manifest invariance}

There are many approaches to supersymmetric field theory, among which superspace formulations hold a prominent place, as they allow for a more geometric understanding of susy. What all these approaches have in common, however, regarding the issue of the on-shell closure of susy transformations on the fields, is the inclusion of auxiliary fields in the theory -- either a finite 
or  infinite number of them \cite{Galperin:2001seg}. This enlarge the field space of the original theory and modifies the susy transformations s.t. the Lagrangian is kept (quasi-)invariant and the (new) susy transformations close a superalgebra off-shell, without implementing the equations of motion. 

\smallskip

Here we adopt a different perspective: Using the DFM, we construct \emph{susy-invariant}, variables -- which one may call susy singlets, to borrow a group-theoretic term -- namely \emph{relational variables} that are invariant under the susy transformations leaving (quasi-)invariant  the bare Lagrangian under consideration. 
 As we will see, applying the dressing procedure to a bare Lagrangian naturally yields a susy-invariant theory; however, one could also choose 
to construct a new theory entirely from scratch, defined in terms of the dressed variables, and quasi-invariant under the residual symmetries.

This process leads to a reshuffling of the d.o.f., susy is effectively reduced, and the need for auxiliary fields disappears, as the issue of on-shell closure of the susy algebra is circumvented. 
That said, the matching of bosonic and fermionic d.o.f. may still pose challenges. In Section \ref{Discussion}, we suggest possible strategies to address this issue. One such approach could involve reintroducing auxiliary fields solely to balance the d.o.f., which can then also be dressed -- since all fields must be dressed when reducing a symmetry.

In the following, we formalize our approach to the challenge. 
The perturbative dressing will be the one we are primarily interested in, particularly because susy transformations in standard supersymmetric field theory are typically given (or derived) at the infinitesimal level. Our approach is in principle applicable to any amount of susy $\mathcal{N}$ and in any spacetime dimension $D$.

\subsection{Susy as an off-shell invariance via the DFM}

We consider a generic supersymmetric field theory with field content denoted by $\upphi$ and Lagrangian $L=L(\upphi)$ quasi-invariant under infinitesimal susy transformations
\begin{align}
\label{generalsusytr}
    \delta_\varepsilon \upphi = f (\varepsilon; \upphi),
\end{align}
where $f$ is a functional expression linear in the susy spinor parameter $\varepsilon=\varepsilon(x)$.
In general, one has
\begin{align}
\label{onshellclosure}
[\delta_\varepsilon,\delta_{\varepsilon'}] \upphi = \delta_{[\varepsilon,\varepsilon']} \upphi +  \bs E (\upphi),
\end{align}
with $\bs E (\upphi)$ the field equations of the theory, the susy transformations closing an algebra only on-shell, i.e. for $\bs E(\upphi)=0$, on the fields $\upphi$. 

Now, suppose that we are able to extract from the theory a \emph{perturbative dressing field} $\upsilon=\upsilon[\upphi]$  
s.t.
\begin{align}
    \delta_\varepsilon \upsilon \approx - \varepsilon .
\end{align}
We can then systematically build \emph{perturbatively dressed fields}
\begin{align}
    \upphi^\upsilon := \upphi + \b\updelta_\upsilon \upphi = \upphi + f(\upsilon;\upphi) .
\end{align}
Those are relational variables, invariant under susy transformations at 1st order. Indeed, we have
\begin{equation}
\begin{aligned}
    \delta_\varepsilon \upphi^\upsilon & = \delta_\varepsilon \upphi + \delta_\varepsilon f (\upsilon;\upphi) \\
    & = f(\varepsilon;\upphi) + f (\delta_\varepsilon \upsilon;\upphi) + \cancelto{\,\text{\tiny{neglect}}}{  \delta_\varepsilon f (\upsilon; \upphi)}\!\!\! \\
    & =  f(\varepsilon;\upphi) - f(\varepsilon;\upphi) = 0 ,
\end{aligned}
\end{equation}
neglecting higher-order terms.

Since all the fields in the theory shall now be dressed in such a way as to be invariant under susy transformations, it follows automatically that 
\begin{align}
\label{issuedefusedoffshell}
[\delta_\varepsilon,\delta_{\varepsilon'}] \upphi^\upsilon = 0 ,
\end{align}
which defuses the issue of the on-shell closure of susy transformations on the fields, which arose when considering only bare variables \eqref{onshellclosure}.

In the presence of residual $\J$-transformations of the 1st kind, the theory is not yet fully relational; these should be reduced in an analogous manner to achieve full relationality. The theory thus becomes a 
$\J$-theory, and the dressed fields still transform under the residual gauge symmetries remaining after susy has been reduced (e.g., Lorentz transformations, bosonic spacetime diffeomorphisms, etc.).

We will now examine the implications for the dynamics of the theory, that is, for supersymmetric Lagrangians.

\subsection{Dressed susy Lagrangians}\label{Dressed susy Lagrangians}

Let us consider the Lagrangian form $L=L(\upphi)$ quasi-invariant under the infinitesimal susy transformations \eqref{generalsusytr}, namely
\begin{align}
    \delta_\varepsilon L(\upphi) = d\beta(\varepsilon;\upphi).
\end{align}
Supposing that there exists the dressing field $\upsilon$ above, 
we exploit the quasi-invariance of $L$ to define, using the DFM rule of thumb, the perturbatively \emph{dressed Lagrangian} as
\begin{equation}
\begin{aligned}
\label{dressed-Lagrangian}
L^\upsilon = L(\upphi^\upsilon) &:=L(\upphi) + \b\updelta_\upsilon L \\
& \,= L(\upphi) + d \beta(\upsilon;\upphi).
\end{aligned}
\end{equation}
The dressed Lagrangian $L^\upsilon$ is  susy-invariant at 1-st order by construction, being a functional of the dressed fields $\upphi^\upsilon$:
\begin{equation}
\begin{aligned}
    \delta_\varepsilon L^\upsilon &= \delta_\varepsilon L + \delta_\varepsilon d\beta (\upsilon;\upphi) \\
    & = \delta_\varepsilon L + d \beta (\delta_\varepsilon \upsilon; \upphi) + \cancelto{\,\text{\tiny{neglect}}}{ d\,  \beta (\upsilon; \delta_\varepsilon\upphi)}\!\!\! \\
    & = d \beta (\varepsilon;\upphi) - d \beta (\varepsilon;\upphi) = 0. 
\end{aligned}
\end{equation}
The dressed field equations $\bs E(\upphi^\upsilon)=0$ are thus susy-invariant at 1st order. They are deterministic, meaning that they uniquely determine the evolution of the relational d.o.f. of the theory.

Observe that, in the case in which the bare Lagrangian is strictly susy-invariant, that is $\beta=0$, then $L(\upphi^\upsilon)= L(\upphi)$. 
In either cases, the field equations $\bs E(\upphi^\upsilon)=0$ for the dressed fields have the \emph{same functional expression} as the field equations for the bare fields, $\bs E(\upphi)=0$.
Let us also mention that in the presence of residual $\J$-transformations of the 1st kind, $L(\upphi^\upsilon)$ is a $\J$-theory. 

We remark that, at this point, one can either continue studying the dressed theory constructed starting from the bare one -- for example, by analyzing its solutions, performing quantization, etc. -- with manifest invariance at hand, or, alternatively, one can construct a new $\J$-theory from scratch using the dressed kinematical field variables obtained by reducing the susy transformations \eqref{generalsusytr}. 

\subsection{Susy as an off-shell invariance in $\mathcal{N}=2$, $D=4$ supergravity via the DFM}

We now give an application of the above to the case of $\mathcal{N}=2$, $D=4$ pure sugra. Let us mention that, actually, in the case of sugra things become a bit more subtle because susy transformations correspond to \emph{diffeomorphisms along the fermionic directions} of \emph{superspace}.
This is made clearer in geometric approaches to susy and sugra in superspace: See for example the so-called \emph{rheonomic} approach \cite{Neeman-Regge1978,Neeman-Regge1978b,Castellani:1991eu} comprehensively reviewed in  \cite{JTF-Ravera2024review}, or more generally the framework of Cartan super-geometry as the mathematical foundation of sugra -- see \cite{JTF-Ravera2024review} and references therein. 
In order to have better geometric control over the procedure to be implemented via the DFM for reducing susy in sugra one should adopt a super-Cartan approach, that is, work with a super-Cartan bundle. 
Moreover, since higher-dimensional, $\mathcal{N}$-extended sugra models typically involve higher-degree forms (antisymmetric tensors with multiple indices), a fully developed relational framework for sugra would require the formal development of higher Cartan supergeometry, 
along with a corresponding higher DFM (in preparation \cite{JTF-Ravera2025higherDFM}).

Here we shall just consider the $\mathcal{N}=2$, $D=4$ pure sugra model as presented in \cite{Castellani:1991eu}, applying the DFM at an elementary field-theoretic level, which is enough to showcase our  approach to the issue of on-shell closure of the susy transformations. 

\smallskip

The field content of the $\mathcal{N}=2$ theory is given by $\upphi^\Sigma=\lbrace \omega^{ab}, V^a, \psi_A, \A_{AB} \rbrace$, with $\omega^{ab}$ the Lorentz spin connection, $V^a$ the vierbein, $\psi_A$ the gravitino 1-form fields, and $\A_{AB}$ the graviphoton (here $\Sigma$ denotes the index of the $\rm{OSp}(4|2)$ adjoint multiplet and $A,B,\ldots =1,2$ are $\rm{SO}(2)$ indices, while $a,b,\ldots=0,1,2,3$). One can introduce the notation $\A^{AB}:= \epsilon^{AB}\A$, with $\epsilon^{AB}=-\epsilon^{BA}$ and $\epsilon^{12}=1$. We shall adopt the same notation and conventions of \cite{Castellani:1991eu}.
The supercurvature 2-forms of the theory are defined as
\begin{equation}
\begin{aligned}
    R^{ab} &:= d \omega^{ab} - {\omega^a}_c \wedge \omega^{cb} + 4 e^2 V^a  V^b + e \b\psi_A \gamma^{ab} \psi_A , \\
    R^a &:= \D V^a - \tfrac{i}{2} \b\psi_A \gamma^a \psi_A = dV^a - {\omega^a}_b V^b - \tfrac{i}{2} \b\psi_A \gamma^a \psi_A , \\
    \rho_A &:= \nabla \psi_A - i e \gamma_a \psi_A V^a , \\
    \F &:= F + \epsilon_{AB} \b\psi_A \psi_B ,
\end{aligned}
\end{equation}
where $\nabla \psi_A:= d\psi_A - \tfrac{1}{4} \omega^{ab} \gamma_{ab} \psi_A + e \epsilon_{AB} A \psi_B $ is the $(\rm{SO}(1,3)\times \rm{SO}(2))$-covariant derivative of $\psi_A$, $F:= dA$, and $e$ is the scale parameter -- proportional to the inverse of the AdS radius -- related to the (negative) cosmological constant by $\Lambda=-48 e^2$.

The geometric superspace Lagrangian reads
\begin{align}
    L  =& R^{ab} V^c V^d \epsilon_{abcd} + 4 \b\rho_A \gamma_5 \gamma_a \psi_A V^a + 2 i \epsilon_{AB} \F \b\psi_A \gamma_5 \psi_B \notag\\
    & - 2 e^2 \epsilon_{abcd} V^a V^b V^c V^d - e \b\psi_A \gamma_{ab} \psi_A V_c V_d \epsilon^{abcd} \notag\\
    & + \b\psi_A \psi_B \b\psi_A \gamma_5 \psi_B + \tfrac{1}{4} F^{ab} V^c V^d \F \epsilon_{abcd} \notag\\
    & - \tfrac{1}{48} F_{ab}F^{ab} V^c V^d V^e V^f \epsilon_{cdef} ,
\end{align}
and is (quasi-)invariant under the following infinitesimal susy transformations of the fields:
\begin{align}
        \delta_\varepsilon V^a  =& i \b\varepsilon_A \gamma^a \psi_A , \\
        \delta_\varepsilon \psi_A  =& \nabla \varepsilon_A + \epsilon_{AB} V^b \left( i  F_{ab} + \tfrac{1}{2}  F^{cd} \gamma_5 \epsilon_{abcd} \right) \gamma^a \varepsilon_B , \notag\\
        \delta_\varepsilon \A  =& 2 \epsilon_{AB} \b\psi_A \varepsilon_B ,\notag \\
        \delta_\varepsilon \omega^{ab}  =& 2 e \b\psi_A \gamma^{ab}\varepsilon_A + \left( i \b\rho_A^{ca} \gamma^b + i \b\rho_A^{cd}\gamma^a - i \b\rho_A^{ab}\gamma^c \right) \varepsilon_A V_c \notag \\
        & + 2 \epsilon_{AB} F^{ab} \b\psi_B \varepsilon_A + i \epsilon^{abcd} \epsilon_{AB} F_{cd} \b\psi_B \gamma_5 \varepsilon_A , \notag
\end{align}
where $\varepsilon_A$ is the $\N=2$ susy parameter and $F_{ab}$ and $\rho_{A|ab}$ are, respectively, the supercovariant field strengths of $\A$ and $\psi_A$. 
We observe that the susy transformation of $\psi_A$ can be formally rewritten as
\begin{align}
\delta_\varepsilon \psi_A = \bigtriangledown \varepsilon_A ,
\end{align}
in terms of a new linear differential operator $\bigtriangledown$ whose action on the susy parameter is defined as
\begin{align}
    \bigtriangledown \varepsilon_A := \D \varepsilon_A + \epsilon_{AB} K \varepsilon_B ,
\end{align}
where
\begin{align}
   K := e \A + V^b \left( i  F_{ab} + \tfrac{1}{2}  F^{cd} \gamma_5 \epsilon_{abcd} \right) \gamma^a . 
\end{align}
Let us now consider the reducible decomposition of $\psi_A$ most commonly mentioned in the susy literature, that is the gamma-trace decomposition
\begin{align}
\label{gammatracedec}
    \psi_{A|\mu}  := {\uprho}_{A|\mu} + \gamma_\mu \chi_A ,
\end{align}
where, in $D=4$ spacetime dimensions, $\chi_A := 1/4 \gamma^{\mu} \psi_{A|\mu}$ is a spin-1/2 field and $\uprho_{A|\mu}$ is s.t. $\gamma^\mu\uprho_{A|\mu} =0$.
We may then consider the dressing functional constraint
\begin{align}
    \gamma^\mu \psi_{A|\mu}^\upsilon = 0
\end{align}
on the \emph{perturbatively dressed variable} 
\begin{align}
    \psi_A^\upsilon := \psi_A + \b\updelta_\upsilon \psi_A ,
\end{align}
and solve it explicitly in terms of the super dressing field $\upsilon_A=\upsilon_A[\psi]$:
\begin{align}
    & \gamma^\mu \psi^\upsilon_{A|\mu} = \gamma^\mu (\psi_{A|\mu} + \bigtriangledown_\mu \upsilon_A) = 0 \\
    &\Rightarrow \quad \upsilon_A = \upsilon_A [\psi] = - \slashed{\bigtriangledown}^{-1} (\gamma^\mu \psi_{A|\mu}) =: - 4 \slashed{\bigtriangledown}^{-1} \chi_A. \notag
\end{align}
We can easily check that $\upsilon_A[\psi]$ properly transforms under susy as a perturbative dressing field (see eq. \eqref{dressingfieldtr}), neglecting higher-order terms. 
Indeed,  
since, as one can easily derive considering the gamma-trace decomposition \eqref{gammatracedec}, $\delta_\varepsilon \chi_A = \frac{1}{4} \slashed{\bigtriangledown} \varepsilon_A$, one has
\begin{align}
    \delta_\epsilon \upsilon_A[\psi]&= 
     - 4 \slashed{\bigtriangledown}\- (\delta_\varepsilon  \chi_A) 
     - 4 \cancelto{\,\text{\tiny{neglect}}}{\delta_\varepsilon  (\slashed{\bigtriangledown}\-) \chi_A}\!\!\!\!\! \notag\\
     &\approx - \varepsilon_A.
\end{align}
So $\upsilon_A[\psi]$ is indeed a (non-local) perturbative dressing field, and we have built the perturbatively (self-)dressed gravitinos
\begin{align}
    \psi^\upsilon_A := \psi_A + \bigtriangledown \upsilon_A[\psi] 
    = \psi_A - 4 \bigtriangledown \slashed{\bigtriangledown}\- \chi_A .
\end{align}
The latter, by construction, identically fulfills $\gamma^{\mu} \psi_{A|\mu}^\upsilon\equiv0$, it is \emph{susy-invariant at 1st order}, $\delta_\varepsilon \psi_A^\upsilon \approx 0$, and carries $12\times 2 = 24$ d.o.f. off-shell.

One shall now also dress the other fields of the theory accordingly. The perturbatively dressed vierbein, graviphoton, and spin connection 1-forms are formally given by
\begin{equation}
\begin{aligned}
    (V^{a})^\upsilon &:= V^a + i \b \upsilon_A[\psi]  \gamma^a \psi_A , \\
    (\A)^\upsilon &:= \A + 2 \epsilon_{AB} \b\psi_A \upsilon_B , \\
    (\omega^{ab})^\upsilon &:= \omega^{ab} + 2 e \b\psi_A \gamma^{ab}\upsilon_A \\
    & + \left( i \b\rho_A^{ca} \gamma^b + i \b\rho_A^{cd}\gamma^a - i \b\rho_A^{ab}\gamma^c \right) \upsilon_A V_c  \\
    & + 2 \epsilon_{AB} F^{ab} \b\psi_B \upsilon_A + i \epsilon^{abcd} \epsilon_{AB} F_{cd} \b\psi_B \gamma_5 \upsilon_A .
\end{aligned}
\end{equation}
All the perturbatively dressed fields above are \emph{relational variables} \cite{JTF-Ravera2024gRGFT}: they represent the physical, susy-invariant relations among the d.o.f. of $\omega^{ab}$, $V^a$, $\A$, and $\psi_A$. 

According to the DFM, as previously discussed, with the dressing fields $\upsilon_A[\psi]$ the Lagrangian 4-form of the dressed theory is given by
\begin{equation}
\begin{aligned}
    L^\upsilon & = L(\omega^\upsilon,V^\upsilon,\A^\upsilon,\psi^\upsilon_A) \\
    & = L(\omega, V, \A, \psi_A) + d\beta(\upsilon_A[\psi] ;\omega, V, \A, \psi_A). 
\end{aligned}
\end{equation}
Let us stress that it is  susy-invariant at 1st order because it is a functional of  the dressed field-theoretic variables $\upphi^\upsilon$, which are susy-invariant at 1st order:
\begin{equation}
\begin{aligned}
    \delta_\varepsilon L^\upsilon &= \delta_\varepsilon L + \delta_\varepsilon d\beta (\upsilon_A[\psi];\omega, V, \A, \psi_A) \\
    & = \delta_\varepsilon L + d \beta (\delta_\varepsilon \upsilon_A[\psi]; \upphi) + \cancelto{\,\text{\tiny{neglect}}}{ d\, \delta_\varepsilon \beta (\upsilon_A[\psi];\upphi)}\!\!\! \\[0.2cm]
    & = d \beta (\varepsilon_A;\omega, V, \A, \psi_A) - d \beta (\varepsilon_A;\omega, V, \A, \psi_A) = 0. 
\end{aligned}
\end{equation}
The dressed field equations $\bs E(\upphi^\upsilon)=0$ have the same functional expression of the bare ones, $\bs E(\upphi)=0$ (given in \cite{Castellani:1991eu}) and they are susy-invariant at 1st order. They are deterministic, meaning that they uniquely determine the evolution of the relational d.o.f. of the theory.
This showcases the relational version of the $\mathcal{N}=2$, $D=4$ pure sugra model, where susy-invariance is implemented at 1st order.

Since all the fields in the model are now dressed in
such a way as to be invariant under susy transformations,
it follows that
\begin{equation}
\begin{aligned}
&[\delta_\varepsilon,\delta_{\varepsilon'}] \,(V^a)^\upsilon = 0 , \\
&[\delta_\varepsilon,\delta_{\varepsilon'}] \,(\psi_A)^\upsilon = 0 , \\
&[\delta_\varepsilon,\delta_{\varepsilon'}] \,(\A)^\upsilon = 0 , \\
&[\delta_\varepsilon,\delta_{\varepsilon'}] \,(\omega^{ab})^\upsilon = 0 , \\
\end{aligned}    
\end{equation}
which in fact circumvents the issue of the on-shell closure of susy
transformations on the (bare) fields.

Observe that the dressed fields still transform under the residual
gauge symmetries of the 1st kind remaining after susy has been reduced, that is under Lorentz transformations, $\rm{SO}(2)$ gauge transformations, and bosonic spacetime diffeomorphisms. Calling the residual gauge group $\J$, we can thus say that the model in which susy has been reduced is thus a $\J$-theory.

\smallskip

Let us finally mention that the same procedure of susy reduction via the DFM can be applied to the $\mathcal{N}=2$ supersymmetric extension of the MacDowell-Mansouri theory presented in \cite{Andrianopoli:2014aqa}, where a boundary term is added to the bulk $\mathcal{N}=2$, $D=4$ sugra theory to restore susy invariance of the action in the presence of a non-trivial boundary of spacetime -- that is, when the fields do not asymptotically vanish at the boundary. 
This situation, whereby  the presence of non-trivial boundaries in $M$ break $\Diff(M)$ (or superdiffeomorphisms, in supersymmetric field theory in superspace) or $\H$-gauge symmetries, also commonly referred to as the \emph{boundary problem}, has the same conceptual structure as a \emph{hole argument}, and can be therefore dissolved by the \emph{point-coincidence argument} -- see, e.g., \cite{JTF-Ravera2024c} and references therein. 
This is technically implemented via the DFM \cite{JTF-Ravera2024gRGFT,JTF-Ravera2025bdy}. The boundary problem dissolves once it is recognised that a \emph{physical spacetime boundaries} are \emph{relationally defined}, and therefore are invariant.

Regardless of this,  observation, and as we have  discussed in Section \ref{Dressed susy Lagrangians}, one can dress the field content as discussed above and consider another Lagrangian functional of the dressed variables for the $\mathcal{N}=2$, $D=4$ sugra theory, such as the supersymmetric MacDowell-Mansouri functional.

\section{Discussion}\label{Discussion}

We have shown that applying the DFM to supersymmetric field theory allows one to circumvent the problem of on-shell closure of susy transformations. 
This leads to a new, manifestly invariant and relational approach to supersymmetric theories, in which susy is reduced. 
Typically, the procedure involves non-local dressing fields, yet the resulting formulation may prove more amenable to quantization, as it achieves manifest invariance without the need for auxiliary fields. 
In doing so, we have focused primarily on point (ii) discussed in the introduction.

As for point (i), the issue of matching bosonic and fermionic d.o.f. off-shell (and of course on-shell) may be addressed by circumventing the traditional no-go theorem of susy \cite{Galperin:2001seg}, which asserts the impossibility of off-shell formulations without an infinite number of auxiliary fields. A viable strategy involves introducing a finite number of auxiliary fields by making use of the so-called (dual formulation of) hidden superalgebras underlying supersymmetric theories in the presence of tensor fields -- namely, higher-degree differential forms. 
Alternative off-shell formulations, potentially involving a finite set of auxiliary fields, could in fact emerge from superspace extensions distinct from the standard harmonic or projective ones. A promising avenue in this direction is indeed offered by the \emph{hidden superalgebra} framework \cite{DAuria:1982uck}, which defines an extension of superspace incorporating not only additional even directions but also new odd directions associated with spin-3/2 fields \cite{DAuria:1982uck,Andrianopoli:2016osu,Penafiel:2017wfr,Andrianopoli:2017itj,Ravera:2018vra,Ravera:2021sly}. Such an extension might allow one to evade the assumptions of the no-go theorem, which traditionally only considers auxiliary spin-1/2 fields, as anticipated in \cite{Andrianopoli:2024ynz}, in particular thinking of (re-)constructing the theory directly in terms of the extra forms involved. Preliminary results supporting this approach can be found in \cite{Concha:2018ywv,Andrianopoli:2021rdk}.

Since these frameworks involve higher-degree differential forms (see also \cite{Ravera:2022vjy}), it is natural to expect that the notion of higher dressing will become relevant in this context \cite{JTF-Ravera2025higherDFM}. Moreover, to construct the full spectrum of higher structures within a superspace formalism, recent developments in the literature -- such as the use of the Poincaré–Hilbert series and the Molien–Weyl formula \cite{Cremonini:2022cdm, Cremonini:2024ddc} -- may offer particularly powerful tools.

Another possibility consists in reducing susy by implementing alternative dressing functional constraints, different from those presented so far for standard susy -- for instance, the one considered in \cite{JTF-Ravera2024ususyDFM}. These allow for a different reduction and rearrangement of bosonic and fermionic d.o.f., already at the purely kinematical level, which may assist in achieving a matching of bosonic and fermionic d.o.f. off-shell. For example, one may think of reducing the off-shell d.o.f. of the extra fermionic 1-form fields appearing in the dual form of the hidden superalgebra to spin-1/2 fields, reducing the symmetry transformations naturally induced by the presence of these objects. A similar argument may also be applied to the (extra) bosonic 1-form fields. 

If, on the other hand, one does not impose or is not interested in requiring the off-shell matching of bosonic and fermionic d.o.f. \cite{Sohnius:1985qm}, it becomes interesting to explore alternative, non-standard approaches to supersymmetric field theory, such as the Matter-Interaction Supergeometric Unification (MISU) proposed in \cite{JTF-Ravera2024ususyDFM,JTF-Ravera2025DFMSusyReview&Misu}.
The latter is close to Berezin's original motivation for the introduction of supergeometry in fundamental physics \cite{Berezin-Marinov1977}, and 
is indifferent to the ultimate empirical status of standard supersymmetric field theory.

\smallskip

{
Irrespective of the chosen formalism, the DFM offers a versatile framework for systematically reducing the symmetries of a theory, including susy. By properly building  a dressing field (whenever possible), one can effectively reduce the symmetry, thus building \emph{manifestly relational}, \emph{automatically invariant} observables. This paves the way for a novel quantization strategy we call \emph{Relational Quantization}, and which we elaborate in a forthcoming paper \cite{JTF-Ravera2025relquant} (see also for preliminary results and examples \cite{JTF-Ravera2024NRrelQM}).}

\begin{acknowledgments}
J.F. is supported by the Austrian Science Fund (FWF), grant \mbox{[P 36542]}, 
and by the Czech Science Foundation (GAČR), grant GA24-10887S.
L.R. is supported by the 
GrIFOS research project, funded by the Ministry of University and Research (MUR, Ministero dell'Università e della Ricerca, Italy), PNRR Young Researchers funding program, MSCA Seal of Excellence (SoE), 
CUP E13C24003600006, ID SOE2024$\_$0000103, of which this paper is part.
\end{acknowledgments}


\bibliography{biblioff.bib}

\end{document}